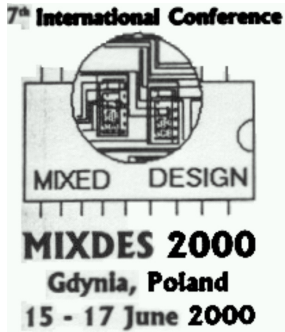

# CONDITIONING OF PIEZORESISTANCE COEFFICIENT EXTRACTION

Z. GNIAZDOWSKI, J. KOSZUR, P. KOWALSKI,
INSTITUTE OF ELECTRON TECHNOLOGY, POLAND

**ABSTRACT**: The necessary and sufficient condition for the piezoresistance coefficient extraction and conditioning of the extraction problem are considered as a problem of the certain matrix *A*. This matrix is implied by the stress distribution on the certain test structure. For the given test structure matrix *A* was calculated and the condition number was estimated. Obtained value of condition number shows that proposed test structure gives well-conditioned matrix *A*. Both the geometrical analysis and numerical estimation of condition number shows that the problem of extraction is well-conditioned and the test structure is properly designed. For the given condition number the error propagation of the input data was considered. For assumed levels of the input data errors, the relative error of piezoresistance coefficient is less than 8%.

## INTRODUCTION

Piezoresistivity is a material property where the bulk resistivity is influenced by the mechanical stresses applied to the material. Monocrystalline silicon has a high piezoresistivity and excellent mechanical properties. These features make silicon particularly suited for the conversion of mechanical deformation into the electrical signal. Therefore, silicon is widely used as a basic material for piezoresistive sensors for mechanical signals such as pressure, flow, force, and acceleration. The requirements for performances of piezoresistive sensors induce the necessity of optimisation of its functional parameters. Modelling is the significant tool in optimisation process. Longitudinal piezoresistance coefficient $\pi_L$ and transversal piezoresistance coefficient $\pi_T$ are necessary for modelling piezoresistors. The method for calculating piezoresistance coefficients $\pi_L$ and $\pi_T$ for homogenous layers is known from literature [1], [2], [3], [4]. This method requires well-defined piezoresistance tensor $\Pi$ dependent on the type of conductivity and doping concentration of the semiconductor [1], [2], [3], [4], [5]. To obtain required accuracy, ion implantation and diffusion is used for fabrication of piezoresistors. In this case modelling with piezoresistance coefficient extracted for uniform doped semiconductor is out of credibility.

For that reason, effective longitudinal and transverse piezoresistance coefficients should be extracted for the given technology. The useful approach for solving this problem was proposed [6]. The adequate test structure is necessary for extraction of piezoresistance coefficients. Apart from this method and the certain test structure we have to investigate some additional problems like conditioning of piezoresistance coefficient extraction.

## METHOD FOR EXTRACTION OF PIEZORESISTANCE COEFFICIENT

For piezoresistor of layout presented in Fig. 1 the resistance $R$ under the stress can be calculated using the following formula:

$$R = R_0 + \rho_0 \pi_L \int_{x_d}^{x_u} \sigma_L(x)dx + \rho_0 \pi_T \int_{x_d}^{x_u} \sigma_T(x)dx \qquad (1)$$

where $\pi_L$ and $\pi_T$ are longitudinal and transverse components of the piezoresistance coefficient respectively, $\sigma_L(x)$ and $\sigma_T(x)$ are corresponding stress components along piezoresistor, $R_0$ is the measured value of piezoresistor without the stress and $\rho_0 = R_0/L$. If we denote:

$$a_L = \int_{x_d}^{x_u} \sigma_L(x)dx \qquad (2a)$$

$$a_T = \int_{x_d}^{x_u} \sigma_T(x)dx \qquad (2b)$$

also $\Delta R = R - R_0$ and $d = \Delta R/\rho_0$, we get:

$$d = a_L \pi_L + a_T \pi_T . \qquad (3)$$

If two different resistors on the test structure are considered, we have the system of two linear equations with two unknown piezoresistance coefficients:

$$\begin{bmatrix} d_1 \\ d_2 \end{bmatrix} = \begin{bmatrix} a_{1L} & a_{1T} \\ a_{2L} & a_{2T} \end{bmatrix} \begin{bmatrix} \pi_L \\ \pi_T \end{bmatrix} . \qquad (4)$$

In matrix notation, this system has the form:

$$d = A\pi \qquad (5)$$

The resolutions of this system for the given technology are values of $\pi_L$ and $\pi_T$. In system (5) vector $d$ is obtained from measurement and matrix $A$ is calculated by Finite Elements Method (FEM) simulation.

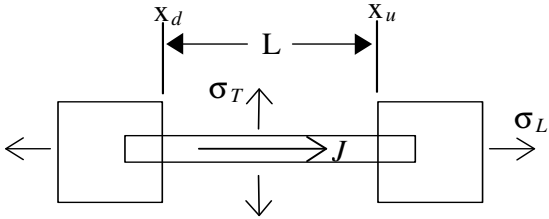

*Fig. 1. Layout of the diffused resistor: L equal to $x_u$-$x_d$ is a length of the resistor, $\sigma_L$ and $\sigma_T$ are stress components and J is a current direction*

## TEST STRUCTURE

As a test structure for extraction of piezoresistance coefficients, silicon pressure sensor is considered (Fig. 2). The structure was fabricated in standard CMOS technology for 5 μm design rule with incorporated anisotropy etching. N-type <100> silicon wafers were applied as a substrate. Four p-type resistors oriented with the [110] crystallographic direction are implanted into the n-type membrane (Fig. 2). The membrane is 20μm thick. Piezoresistors are connected into the Wheatstone bridge.

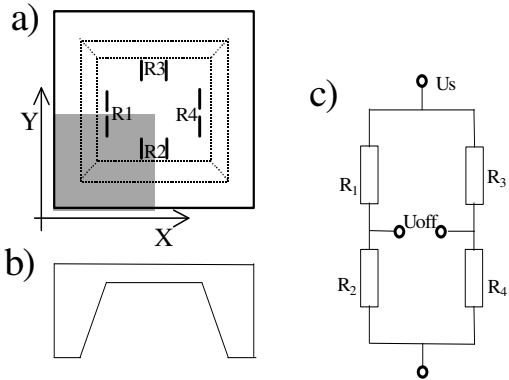

*Fig. 2. The silicon pressure sensor scheme: a) top view of the structure, b) cross-section, c) connection between resistors*

The FEM simulation of the sensor was performed for 100-kPa pressure applied to the membrane. The sensor chip is symmetrical therefore only a quarter of the structure (the grey area on Fig. 2) was modelled. This fact can be taken into consideration in formula (1), without loss of generality. The exact distributions of the stress components $\sigma_L(x)$ and $\sigma_T(x)$ along piezoresistors are extracted from the simulation. Points depicted in Fig. 3 and Fig. 4 represent longitudinal and transverse component of the stress along piezoresistors $R_1$ and $R_2$ calculated using SAMCEF system [7].

The distributions of the stress components across the piezoresistors are substituted by the second order polynomials obtained using the regression analysis method. These polynomials are depicted in Fig. 3 by lines. They are integrated in formula (2), to get the matrix *A*.

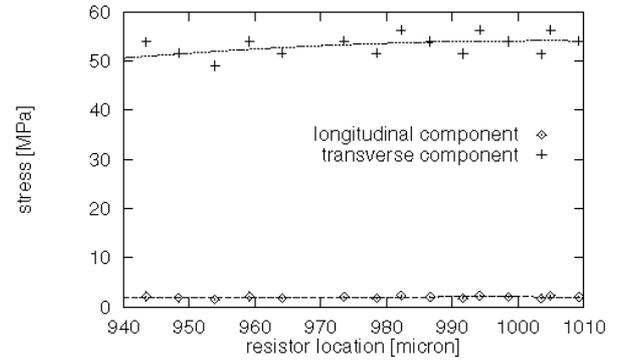

*Fig. 3. Distribution of the stress along half of resistor $R_1$ (located on a grey area in Fig. 2). Points represent the results of FEM modelling. Lines depict second order polynomial models.*

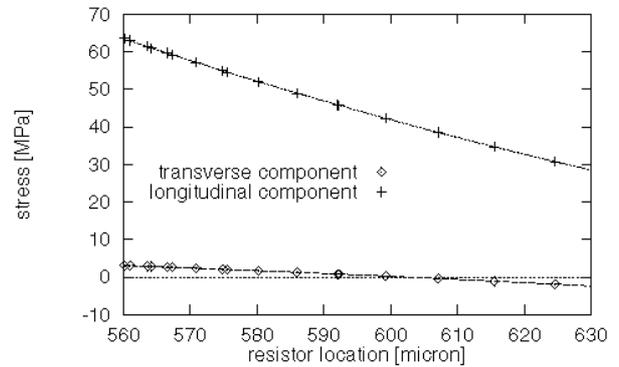

*Fig. 4. Distribution of the stress along half of resistor $R_2$ (located on a grey area in Fig. 2). Points represent the results of FEM modelling. Lines depict second order polynomial models.*

## PROBLEM DEFINITION

Solving of system (5) is required for extraction of piezoresistance coefficient. The resolution of the system is determined by matrix *A*. Especially, necessary and sufficient conditions for solution of this system and error propagation factor are dependent on this matrix. Matrix *A* is implied by the stress distribution on the test structure. Therefore, if system (5) can be resolved with acceptable level of error propagation, it means that the structure was properly designed. We should answer on two questions:
1) Is the problem of extraction resolvable?
2) What is the quality of resolution?

Firstly, necessary and sufficient condition for extraction of piezoresistance coefficient will be considered [8]. If necessary and sufficient condition of the problem is fulfilled, we have to estimate the factor of error propagation. In numerical analysis, this factor is known as a condition number. Condition number is the largest factor by which a relative error in input data can be multiplied when propagate into a relative error in $\pi$. For the given condition number we will estimate errors of the extraction with assumptions about some errors of input data.

## NECESSARY AND SUFFICIENT CONDITION

Necessary and sufficient condition for solving system (5) is:

$$\det(A) \neq 0, \qquad (6)$$

where $det(A)$ denotes determinant of matrix $A$. Formula (3) presents straight line on a plane $\pi_L \times \pi_T$. Coefficients $a_L$ and $a_T$ establish vector $[a_L, a_T]$ normal to this straight line. Condition (6) is equivalent to the linear independence of vectors $a_1 = [a_{1L}, a_{1T}]$ and $a_2 = [a_{2L}, a_{2T}]$. Both vectors $a_1$ and $a_2$ are determined by the arrangement of the test structure. For the given test structure the interaction between this vectors can be interpreted in two manners using geometrical and physical considerations.

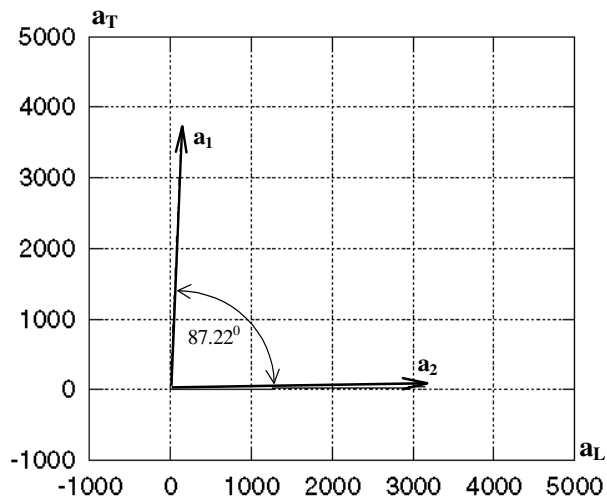

*Fig. 5. Vectors $a_1$ and $a_2$ on a plane $a_L$ versus $a_T$ - both vectors are calculated for location of resistors $R_1$ and $R_2$ on test structure, respectively.*

1. Geometrical consideration: From (2) we have $a_1 = [138.86, 3715.05]$ and $a_2 = [3145.12, 34.84]$. Fig. 5 demonstrates the location of both vectors on a plane $a_L$ versus $a_T$. We consider the inner product of vectors $a_1$ and $a_2$:

   $$a_1 \cdot a_2 = a_{1L} a_{2L} + a_{1T} a_{2T} = |a_1| \cdot |a_2| \cdot \cos(a_1, a_2) \qquad (7)$$

   Calculated value of the angle between both vectors is equal to 87.22 degrees. That means, vectors $a_1$ and $a_2$ are not linearly dependent. These vectors are nearly orthogonal and the matrix $A$ is nearly diagonal.

2. Physical consideration: The transverse component of the stress is predominant component in vector $a_1$ while longitudinal component is relatively small. It means that in resistor $R_1$ $\sigma_T(x) \gg \sigma_L(x)$. On the other hand, longitudinal component is predominant component in vector $a_2$ and transverse component is relatively small. It means that in resistor $R_2$ $\sigma_L(x) \gg \sigma_T(x)$.

Geometrical and physical analysis shows that matrix $A$ has a good property concerning condition (6). This fact is equivalent to the good attribute of the test structure. These good features result from the stress distribution in resistors $R_1$ and $R_1$.

## ERROR PROPAGATION

The problem of error propagation in solving extraction problem has been treated. Assuming no computational error in the solution process but initial errors in matrix $A$ and vector $d$ only, we desire to generate as little error as possible. Our problem is: what will be the error in vector $\pi$ for the given errors in matrix $A$ and vector $d$? For solving this problem, we use estimations well known in numerical analysis [9]. For this purpose, we define the condition number of problem (5):

$$cond(A) = \|A\| \cdot \|A^{-1}\|, \qquad (8)$$

where $\|x\| = \max_{1 \leq i \leq n} |x_i|$ is the Tchebycheff norm of the vector $x$ and $\|A\| = \max_{1 \leq i \leq n} \sum_{j=1}^{n} |a_{ij}|$ is the corresponding matrix norm depends on above mentioned vector norm. In general, the value of condition number is greater than one. Well-conditioned problems have condition numbers between 1 and 10. They propagate relative errors by a factor no larger than 10. Ill-conditioned problem have condition numbers greater than 100. For our test structure, the estimated conditioning number $cond(A)$ is equal to 1.24, therefore our test structure is well-conditioned. This result is consistent with presented above geometrical and physical interpretation. Now we can consider the relative error of $\pi$ assuming errors in matrix $A$ and vector $d$. Relative error of matrix $A$ results from error of definition of the thickness of the membrane of the test structure. Simulation is performed for assumed thickness of the membrane. The assumed thickness can differ from the real thickness resulted from random disturbances in the technological process. Therefore, the matrix $A$ used for solving system (5) can be charged by an error. For our consideration, we assume relative error $\|\Delta A\|/\|A\|$ less than 5.0%. Error of vector $d$ is consisted of three components:

- error of estimation of the length of piezoresistor, because of technological process;
- errors of resistance $R$ and $R_0$ measurement.

For our consideration, we assume that summary relative error $\|\Delta d\|/\|d\|$ is less than 1.0%.

If $cond(A) \cdot \|\Delta A\|/\|A\| < 1$, the relationship between their relative errors and relative error of $\pi$ can be estimated by the formula:

$$\frac{\|\Delta \pi\|}{\|\pi\|} \leq \frac{cond(A)}{1 - cond(A) \cdot \|\Delta A\|/\|A\|} \left( \frac{\|\Delta d\|}{\|d\|} + \frac{\|\Delta A\|}{\|A\|} \right). \qquad (9)$$

From here, we have total relative error $\|\Delta \pi\|/\|\pi\|$ less than 7.95%. Now we can consider relative error of vector $\pi$ assuming two specific cases:

1. If matrix *A* has no error and *d* has error *Δd*, the relative error of $\pi$ is estimated by inequality:

$$\frac{\|\Delta\pi\|}{\|\pi\|} \leq cond(A)\frac{\|\Delta d\|}{\|d\|}. \quad (10)$$

For assumed above relative error of vector *d*, estimated relative error of piezoresistance coefficients $\|\Delta\pi\|/\|\pi\|$ is less than 1.24%.

2. If *d* is error free but matrix *A* has error *ΔA*, we can use the estimation:

$$\frac{\|\Delta\pi\|}{\|\pi\|} \leq \frac{cond(A)\cdot\|\Delta A\|/\|A\|}{1-cond(A)\cdot\|\Delta A\|/\|A\|}. \quad (11)$$

For that reason, the relative error $\|\Delta\pi\|/\|\pi\|$ is less than 6.61%. If $cond(A)\cdot\|\Delta A\|/\|A\| \ll 1$, inequality (11) proceed to:

$$\frac{\|\Delta\pi\|}{\|\pi\|} \leq cond(A)\frac{\|\Delta A\|}{\|A\|}. \quad (12)$$

On the other hand, denoting $\pi^* = \pi + \Delta\pi$ we can write:

$$\frac{\|\Delta\pi\|}{\|\pi^*\|} \leq cond(A)\frac{\|\Delta A\|}{\|A\|}, \quad (13)$$

For assumptions like above, the relative error $\|\Delta\pi\|/\|\pi^*\|$ is less than 6.2%. If $cond(A)\cdot\|\Delta A\|/\|A\| \ll 1$, in (13) we can use $\pi$, instead of $\pi^*$ and inequality (13) becomes to formula (12).

## CONCLUSIONS

The model of piezoresistivity can be tuned to the technological process by extraction of the proper piezoresistance coefficients. Necessary and sufficient condition for extraction problem was considered as a problem of the certain matrix *A*. This matrix is implied by the stress distribution on the certain test structure. For the given test structure the stress distribution implies good matrix A regard to formula (6). For this test structure, the condition number was estimated. Obtained value of condition number shows that proposed test structure gives well-conditioned matrix *A*. Both the geometrical analysis of vectors $a_1$ and $a_2$ and numerical estimation of condition number shows that the problem of extraction is well-conditioned. Suitable vectors are near perpendicular and condition number is near one. For the given condition number the error propagation of the input data was considered. For assumed levels of the input data errors, the relative error of piezoresistance coefficient is less than 8%.

## ACKNOWLEDGEMENTS

The authors would like to express their gratitude to Mr. Jerzy Weremczuk for theirs constructive suggestions. This work was partly supported by the project No. 8T11B03116 from the Polish State Committee for Scientific Research (KBN).

## THE AUTHORS

Dr. Zenon Gniazdowski, Mr Paweł Kowalski and Mr Jan Koszur are with the Institute of Electron Technology, Al. Lotnikow 32/46, 02-668 Warsaw, Poland.
E-mail: gniazd@ite.waw.pl

## REFERENCES

[1] W. Gopel, J. Hesse, J.N. Zemel, "Sensors. A Comprehensive Survey". Vol. 7. "Mechanical Sensors". Edited by H.H. Bau, N.F. de Rooij, B. Kloeck, VCH, Weinheim 1994
[2] O.N. Tufte, D. Long, "Recent Developments in Semiconductor Piezoresistive Devices", Solid-State Electronics, Pergamon Press 1963. Vol. 6, pp. 323-338
[3] Y. Kanda, "Piezoresistance effect of silicon", Sensors and Actuators A, Vol. 28 (1991) 83-91
[4] Y. Kanda, "A Graphical Representation of the Piezoresistance Coefficients in Silicon", IEEE Trans. on Electron Devices, Vol. ED-29, No. 1, Jan. 1982
[5] O.N. Tufte, E.L. Stelzer, "Piezoresistive Properties of Silicon Diffused Layers", J. Appl. Phys. Vol. 34. No. 2, 313-318, Feb. 1963
[6] Z. Gniazdowski, P. Kowalski, "Practical approach to extraction of piezoresistance coefficient", Sensors and Actuators A, Vol. 68, 1998, pp. 329-332
[7] SAMCEF Users Manuals, Samtech, Liege 1996
[8] Z. Gniazdowski, J. Koszur, B. Latecki, "Methodology for modelling of piezoresistors", submitted to IEEE Design and Test of Computers
[9] S.M. Pizer, "Numerical computing and Mathematical Analysis". Science Research Associates, Inc. 1975